\documentclass[prd,english,12pt,nofootinbib]{revtex4}

\usepackage{amssymb}
\usepackage[dvips]{graphicx}

\usepackage[T1]{fontenc}
\usepackage[latin1]{inputenc}

\makeatletter

\usepackage{babel}
\makeatother

\begin{document}

\title{Spontaneous symmetry breaking in a \\ two-doublet lattice Higgs model}

\author{Randy Lewis}
\affiliation{Department of Physics and Astronomy, York University,
Toronto, Ontario, Canada M3J 1P3}

\author{R. M. Woloshyn}
\affiliation{TRIUMF, 4004 Wesbrook Mall, Vancouver, British Columbia, Canada V6T 2A3}

\begin{abstract}

An SU(2) lattice gauge theory with two doublets of complex scalar fields is 
considered. All continuous symmetries are identified and, using the 
nonperturbative methods of lattice field theory, the phase diagram is
mapped out by direct numerical simulation. Two-doublet models contain
phase transitions that separate qualitatively distinct regions of the 
parameter space. In some regions global symmetries are spontaneously
broken. For some special choices of the model parameters, the symmetry-breaking
order parameter is calculated. The pattern of symmetry breaking is 
verified further through observation of Goldstone bosons.
\end{abstract}

\maketitle

\section{Motivation}

The Higgs mechanism lies at the heart of the standard model of electroweak
interactions, where it is implemented efficiently through a single SU(2)
doublet of scalar fields.  
The scalar doublet's mass-squared term is chosen to be negative and 
analysis of small fluctuations around the minimum of the classical scalar
potential leads to the conclusion that the weak gauge bosons acquire 
appropriate masses.
This standard model is generally viewed as the low-energy
effective theory for something more complete, and in many extensions 
multiple scalar doublets appear.
An old, but still useful, review of the Higgs mechanism where multiple
doublets participate may be found in \cite{Sher:1988mj}.

A different specific implementation of two scalar doublets is 
the inert doublet model of \cite{Barbieri:2006dq}(see also \cite{Ma:2006km}).
Instead of taking both scalar doublets to have
nonzero vacuum expectation values (vevs), the inert doublet model assumes 
that there is a phase in which one doublet has a vanishing vev while the 
other does not. This provides some additional motivation for examining
the possible phases of two-doublet models in a more general way.

In this work we use lattice field theory to study some features of 
the Higgs model with two doublets.
Lattice field theory provides a nonperturbative method for studying
non-Abelian gauge theories. An important difference of the lattice formulation 
from the continuum is the use of the gauge field link variable,
which, taking values in the gauge group, allows calculations to be
done without gauge fixing\cite{Balian:1975}. This has the immediate 
consequences that expectation values of non-gauge-invariant operators
vanish and that the gauge symmetry can not be 
spontaneously broken\cite{Elitzur:1975im,Angelis:1978}.
It also implies that physical states of the system are gauge-invariant
composite objects\footnote{A view of the electroweak theory along 
these lines has been espoused by Fr\"ohlich {\em et al.}\cite{Frohlich:1981}.}.
These are features shared by lattice Higgs models and lattice QCD.

Soon after lattice field theory was developed it was applied to Higgs
models\cite{Fradkin:1978dv,Osterwalder:1977pc,Lang:1981qg,Seiler:1982pw,
Kuhnelt:1983mw}.
Lattice simulations of the one-doublet model were carried out 
extensively in the 1980s and early 1990s with applications to the study
of the phase diagram and basic particle spectrum\cite{Montvay:1984wy,
Jersak:1985nf,Evertz:1985fc,Gerdt:1984ft,Langguth:1985dr,Montvay:1985nk,
Evertz:1989hb}, bounds on the scalar (Higgs) mass\cite{Hasenfratz:1987uc}
and the study of
the electroweak finite-temperature phase transition\cite{Jansen:1995yg,
Rummukainen:1996sx,Laine:1998jb,Fodor:1999at}. 
In addition, bounds on the Higgs boson mass have been obtained from
simulations with Higgs-Yukawa theories that omit all gauge
interactions\cite{Bhanot:1990ai,Holland:2003jr,Holland:2004sd,Shigemitsu:1991tc,
De:1991me,Giedt:2007qg,Fodor:2007fn,GerholdJansen}.

An early observation\cite{Fradkin:1978dv,Seiler:1982pw} 
was that the SU(2) lattice Higgs model with a single doublet of scalar fields
in the fundamental representation should actually have only a 
single phase. There are regions in parameter space, 
sometimes named the confinement region
and the Higgs region, which have a qualitatively different mass spectrum.
In most of the parameter space these regions are separated by a phase 
transition. However, there is a corner of parameter space where the transition 
disappears and through which the confined and Higgs regions can be analytically
connected. There is no (local) order parameter and no broken
symmetry to distinguish the two regions. 

With regard to the spectrum, the low-lying states of the one-doublet
SU(2) lattice
Higgs model consist of a scalar singlet and a triplet of vector bosons.
These states are massive in all regions of the parameter space. Note
that the Goldstone bosons which emerge in an intermediate stage of
the standard perturbative treatment of the Higgs mechanism and which
are subsequently absorbed into the longitudinal components of the
massive vector bosons do not appear in the nonperturbative lattice
calculation.

Dramatic qualitative changes may occur when additional scalar doublets are
present in the theory.  In particular, there are two regions of the phase
diagram which are now completely separated by a phase transition throughout
parameter space\cite{Wurtz:2009gf}.  One might expect that these phases are
distinguished by having different global symmetries, and if so, there will be
corresponding order parameters. There may be regions of the parameter
space where global symmetries are spontaneously broken and Goldstone
bosons are present in the spectrum of physical states.
In the present work, we study a gauge theory with two scalar doublets
using numerical lattice simulations in which this scenario is realized.

The lattice action is defined in Sec.~\ref{sec_actionsymmetries} and 
its continuous symmetries are discussed.
Section~\ref{sec_phases} presents the 
numerical simulations used to determine vacuum
expectation values that produce a map of the phase diagram of this two-doublet
lattice Higgs model.
Section~\ref{sec_breakingsGoldstones} describes the methods used to search for
spontaneous symmetry breaking. The symmetry-breaking
order parameter is calculated and
the Goldstone bosons that accompany each broken generator are identified.
Section~\ref{sec_conclusion} contains a summary.

\section{Lattice action and symmetries}\label{sec_actionsymmetries}

This study is based on an action for an SU(2)
gauge theory with two complex scalar doublets where each doublet has its own
global SU(2) symmetry\footnote{This restricts the form of the allowed 
quartic coupling terms. Terms allowed by gauge symmetry but which break
the generic SU(2)$\times$SU(2) global symmetry are excluded here although they
may be present in phenomenological applications\cite{Sher:1988mj}.}.  
On a spacetime lattice, the action can be written as
\begin{equation}\label{action}
S = \sum_x\bigg({\cal L}_g[U] + {\cal L}_1[\Phi_1,U] + {\cal L}_2[\Phi_2,U]
     + {\cal L}_{12}[\Phi_1,\Phi_2]\bigg) \,,
\end{equation}
where
\begin{eqnarray}
{\cal L}_g[U] &=& \frac{\beta}{2}
      \sum_{\mu=1}^4\sum_{\nu=1}^4\left(1-\frac{1}{2}
      {\rm Tr}\left[U_\mu(x)U_\nu(x+\mu)U^\dagger_\mu(x+\nu)U^\dagger_\nu(x)
      \right]\right) \,, \\
{\cal L}_n[\Phi_n,U] &=& \Phi_n^\dagger(x)\Phi_n(x)
      +\lambda_n\bigg(\Phi_n^\dagger(x)\Phi_n(x)-1\bigg)^2 \nonumber \\
 && - \kappa_n\sum_{\mu=1}^4
      \bigg(\Phi_n^\dagger(x+\mu)U_\mu^\dagger(x)\Phi_n(x)
      +\Phi_n^\dagger(x)U_\mu(x)\Phi_n(x+\mu)\bigg) \,, \\
{\cal L}_{12}[\Phi_1,\Phi_2]
 &=& 2\lambda_{12}\Phi_1^\dagger(x)\Phi_1(x)\Phi_2^\dagger(x)\Phi_2(x) \,.
\end{eqnarray}
$U_\mu(x)$ is the gauge field and $\Phi_n(x)$ is a complex scalar doublet.
Notice that the couplings $\lambda_1$ and $\lambda_2$
multiply more than just quartic terms,
in contrast to common practice in the continuum.
Likewise the normalization of the scalar fields $\Phi_n(x)$ follows conventions
of lattice field theory rather than continuum conventions, and thus we show
hopping parameters $\kappa_n$ instead of quadratic coefficients $\mu_n^2$.
The classical relationship between the lattice and continuum
notations may be found in \cite{Kuhnelt:1983mw}.
All parameters and fields in the lattice action are dimensionless.

The 4 degrees of freedom in a complex doublet,
\begin{equation}
\Phi_n(x) = \left(\begin{array}{c} a(x)+ib(x) \\
            c(x)+id(x) \end{array}\right) \,,
\end{equation}
can also be expressed as a matrix,
\begin{equation}
\varphi_n(x) = \left(\begin{array}{rr}
               c(x)-id(x) &~~~~ a(x)+ib(x) \\
              -a(x)+ib(x) & c(x)+id(x)
               \end{array}\right) \,,
\end{equation}
which is a more convenient notation in some contexts.
In this notation, the scalar terms in the Lagrangian become
\begin{eqnarray}
{\cal L}_n[\varphi_n,U]
 &=& \frac{1}{2}{\rm Tr}\left[\varphi_n^\dagger(x)\varphi_n(x)\right]
     +\lambda_n\bigg(\frac{1}{2}{\rm Tr}\left[\varphi_n^\dagger(x)\varphi_n(x)
     \right]-1\bigg)^2 \nonumber \\
  && - \kappa_n\sum_{\mu=1}^4
     {\rm Tr}\left[\varphi_n^\dagger(x)U_\mu^\dagger(x)\varphi_n(x+\mu)\right]
     \,, \\
{\cal L}_{12}[\varphi_1,\varphi_2]
 &=& \frac{\lambda_{12}}{2}{\rm Tr}\left[\varphi_1^\dagger(x)\varphi_1(x)\right]
     {\rm Tr}\left[\varphi_2^\dagger(x)\varphi_2(x)\right] \,.
\end{eqnarray}
This action has one local continuous symmetry,
namely the gauge symmetry defined by
\begin{eqnarray}
U_\mu(x) &\to& R_g(x)U_\mu(x)R_g^\dagger(x+\mu) \,, \\
\Phi_1(x) &\to& R_g(x)\Phi_1(x) \,, \\
\Phi_2(x) &\to& R_g(x)\Phi_2(x) \,,
\end{eqnarray}
where $R_g(x)$ is an element of SU(2).
The action has two global continuous symmetries, one for each scalar doublet,
which will be referred to as intradoublet symmetries.  They are defined by
\begin{eqnarray}
\varphi_1(x) &\to& \varphi_1(x)R_1 \,, \\
\varphi_2(x) &\to& \varphi_2(x)R_2 \,,
\end{eqnarray}
where $R_1$ and $R_2$ are elements of SU(2).
Finally, the action acquires an additional global continuous symmetry in the
special case of ($\kappa_1$=$\kappa_2$, $\lambda_1$=$\lambda_2$=$\lambda_{12}$).
This additional symmetry will
be called the interdoublet symmetry, and it is defined by
\begin{equation}
\left(\begin{array}{l} \Phi_1(x) \\ \Phi_2(x)\end{array}\right)
~~\to~~ R_{12}\left(\begin{array}{l}\Phi_1(x) \\ \Phi_2(x)\end{array}\right)\,,
\end{equation}
where $R_{12}$ is an element of U(2).

It is important to understand the intricate connection between the
interdoublet and intradoublet symmetries.  To elucidate this connection,
use the explicit form
\begin{eqnarray}
R_n = \left(\begin{array}{rr} e^{-i\alpha_n}\cos\gamma_n
 & e^{i\beta_n}\sin\gamma_n \\ -e^{-i\beta_n}\sin\gamma_n
 & e^{i\alpha_n}\cos\gamma_n \end{array}\right)
\end{eqnarray}
for $n=1$ or 2 which gives
\begin{eqnarray}
\Phi_n(x) &\to&
\cos\gamma_ne^{i\alpha_n}\Phi_n(x) + \sin\gamma_ne^{i\beta_n}\Phi_{cn}(x) \,,
\label{Phi} \\
\Phi_{cn}(x) &\to&
\cos\gamma_ne^{-i\alpha_n}\Phi_{cn}(x) - \sin\gamma_ne^{-i\beta_n}\Phi_n(x) \,,
\label{Phic}
\end{eqnarray}
where $\Phi_{cn}(x)\equiv i\tau_2\Phi_n^*(x)$.
The interdoublet symmetry can be parametrized in a similar fashion, but it is
convenient to have it act on
$\left(\begin{array}{l} \Phi_1(x) \\ \Phi_{c2}(x)\end{array}\right)$
rather than on
$\left(\begin{array}{l} \Phi_1(x) \\ \Phi_2(x)\end{array}\right)$, giving
\begin{equation}\label{R12}
\left(\begin{array}{l} \Phi_1(x) \\ \Phi_{c2}(x)\end{array}\right)
\to
\left(\begin{array}{rr} e^{i\delta_{12}}e^{-i\alpha_{12}}\cos\gamma_{12}
& e^{i\delta_{12}}e^{i\beta_{12}}\sin\gamma_{12} \\
 -e^{i\delta_{12}}e^{-i\beta_{12}}\sin\gamma_{12}
& e^{i\delta_{12}}e^{i\alpha_{12}}\cos\gamma_{12} \end{array}\right)
\left(\begin{array}{l} \Phi_1(x) \\ \Phi_{c2}(x)\end{array}\right) \,.
\end{equation}
Now we can combine Eqs.~(\ref{Phi}), (\ref{Phic}) and (\ref{R12}) to
write down the transformation of our scalar fields under
an arbitrary global transformation:
\begin{eqnarray}
\Phi_1 &\to&
  \cos\gamma_{12}e^{i(\delta_{12}-\alpha_{12})}
                           \left(\cos\gamma_1e^{i\alpha_1}\Phi_1
                                +\sin\gamma_1e^{i\beta_1}\Phi_{c1}\right)
  \nonumber \\ &&
+ \sin\gamma_{12}e^{i(\delta_{12}+\beta_{12})}
                           \left(\cos\gamma_2e^{-i\alpha_2}\Phi_{c2}
                                -\sin\gamma_2e^{-i\beta_2}\Phi_2\right) \,,
\label{TenParametersA} \\
\Phi_{c2} &\to&
  \cos\gamma_{12}e^{i(\delta_{12}+\alpha_{12})}
                           \left(\cos\gamma_2e^{-i\alpha_2}\Phi_{c2}
                                -\sin\gamma_2e^{-i\beta_2}\Phi_2\right)
  \nonumber \\ &&
- \sin\gamma_{12}e^{i(\delta_{12}-\beta_{12})}
                           \left(\cos\gamma_1e^{i\alpha_1}\Phi_1
                                +\sin\gamma_1e^{i\beta_1}\Phi_{c1}\right) \,.
\label{TenParametersB}
\end{eqnarray}
Finally we notice that two of the ten parameters
(i.e.\ the $\alpha_i$, $\beta_i$,
$\gamma_i$ and $\delta_i$) are redundant.  Let us choose the eight independent
parameters to be $\gamma_1$, $\gamma_2$, $\gamma_{12}$,
\begin{eqnarray}
\rho_1 &\equiv& \alpha_1 + \delta_{12}
                         - \left(\frac{\alpha_{12}+\beta_{12}}{2}\right) \,, \\
\omega_1 &\equiv& \beta_1 + \delta_{12}
                        - \left(\frac{\alpha_{12}+\beta_{12}}{2}\right) \,, \\
\rho_2 &\equiv& \alpha_2 - \delta_{12} - \left(\frac{\alpha_{12}+\beta_{12}}{2}
                         \right) \,, \\
\omega_2 &\equiv& \beta_2 - \delta_{12} - \left(\frac{\alpha_{12}+\beta_{12}}{2}
                         \right) \,, \\
\theta &\equiv& \frac{\beta_{12}-\alpha_{12}}{2} \,.
\end{eqnarray}
When expressed in terms of the new parameters, Eqs.~(\ref{TenParametersA})
and (\ref{TenParametersB}) become
\begin{eqnarray}
\Phi_j &\to& e^{i\theta}\left[
  \cos\gamma_{12}\left(\cos\gamma_je^{i\rho_j}\Phi_j
                 +\sin\gamma_je^{i\omega_j}\Phi_{cj}\right)
+ \sin\gamma_{12}\left(\cos\gamma_ke^{-i\rho_k}\Phi_{ck}
                 -\sin\gamma_ke^{-i\omega_k}\Phi_k\right)\right] \,,
  \nonumber \\ \label{generalPhiTransformation} \\
\Phi_{cj} &\to& e^{-i\theta}\left[
  \cos\gamma_{12}\left(\cos\gamma_je^{-i\rho_j}\Phi_{cj}
                 -\sin\gamma_je^{-i\omega_j}\Phi_j\right)
- \sin\gamma_{12}\left(\cos\gamma_ke^{i\rho_k}\Phi_k
                 +\sin\gamma_ke^{i\omega_k}\Phi_{ck}\right)\right] \,,
  \nonumber \\ \label{generalPhicTransformation}
\end{eqnarray}
where $(j,k)$=(1,2) or (2,1).
We now recognize the continuous global symmetries as
\begin{equation}
{\rm SU(2)}\times{\rm SU(2)}\times{\rm U(1)}\times{\rm U(1)}
~~~~~{\rm if}~ (\kappa_1=\kappa_2,\lambda_1=\lambda_2=\lambda_{12})
{\rm ~is~true,}
\end{equation}
where the parameters of the four factors are respectively
$(\rho_1,\omega_1,\gamma_1)$, $(\rho_2,\omega_2,\gamma_2)$,
$\gamma_{12}$ and $\theta$.
Of course whenever ($\kappa_1$=$\kappa_2$,
$\lambda_1$=$\lambda_2$=$\lambda_{12}$) is not valid, the continuous
global symmetries are just the intradoublet ones,
\begin{equation}
{\rm SU(2)}\times{\rm SU(2)}
~~~~~{\rm if}~ (\kappa_1=\kappa_2,\lambda_1=\lambda_2=\lambda_{12})
{\rm ~is~not~true,}
\end{equation}
which amounts to choosing $\gamma_{12}=0$.  The parameter $\theta$ is then
redundant, so neither of the U(1) symmetries remains whenever
($\kappa_1$=$\kappa_2$, $\lambda_1$=$\lambda_2$=$\lambda_{12}$) is not valid.

To conclude this section, return to the defining action of Eq.~(\ref{action})
and consider the special case of fixed-length scalar fields,
\begin{equation}
\Phi_1^\dagger(x)\Phi_1(x) = \Phi_2^\dagger(x)\Phi_2(x) = 1 \,.
\end{equation}
In this limit, the theory is independent of parameters $\lambda_1$, $\lambda_2$
and $\lambda_{12}$.  With only the $\kappa_i$ parameters remaining, the
fixed-length theory bears a notable resemblance to QCD-like theories and has
been studied in some detail\cite{Langguth:1985eu,Campos:1997dc,Bonati:2009pf}.

\section{The phase diagram}\label{sec_phases}

In numerical simulations, each scalar or gauge field is evaluated using
a combination of heatbath and over-relaxation updates
combined with an accept-reject step that accounts for
non-Gaussian terms in the action.  The algorithm contains a parameter that
is tuned to produce a good acceptance rate.  Details of the algorithm can be
found in \cite{Wurtz:2009gf}; for more extensive discussions of algorithms
see \cite{Bunk:1994xs,Fodor:1994ih,Fodor:1994sj}.

\begin{figure}
\includegraphics[width=15cm,angle=270,trim=20 350 80 50,clip=true]{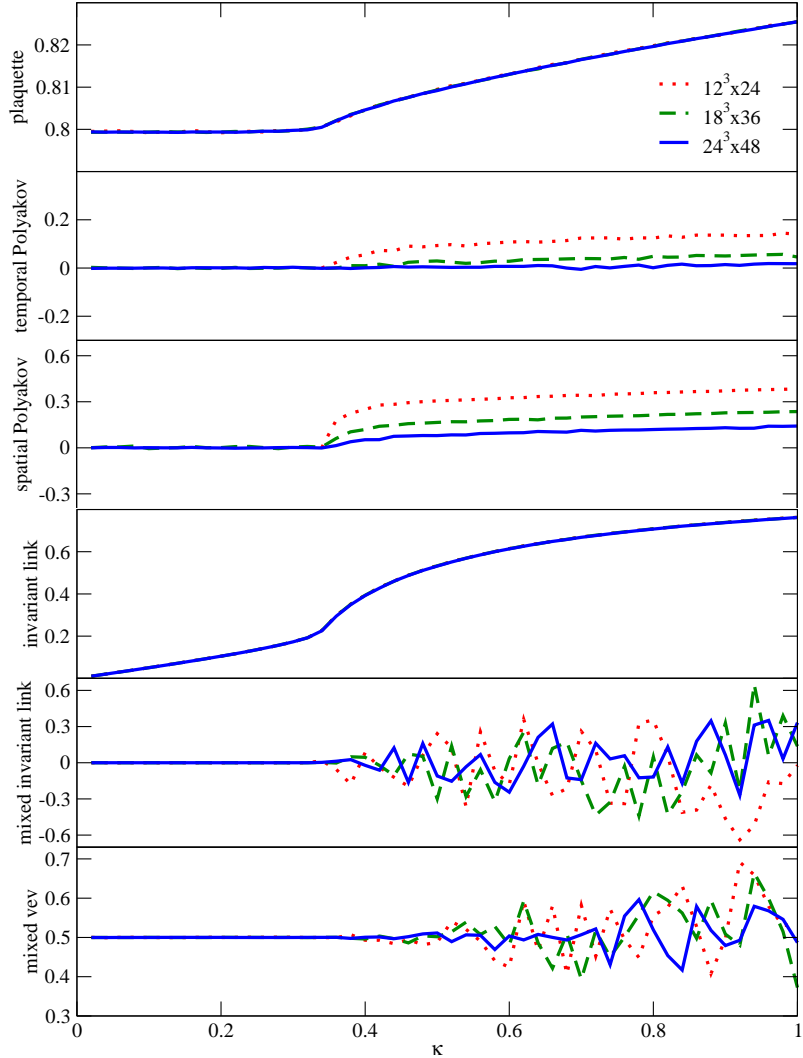}
\caption{Indications of the phase transition from a variety of observables
         for the fixed-length theory at $\beta=4.0$.
         Three different lattice sizes are shown.}
\label{fig_beta4}
\end{figure}

\begin{figure}
\includegraphics[width=15cm,angle=270,trim=20 350 80 50,clip=true]{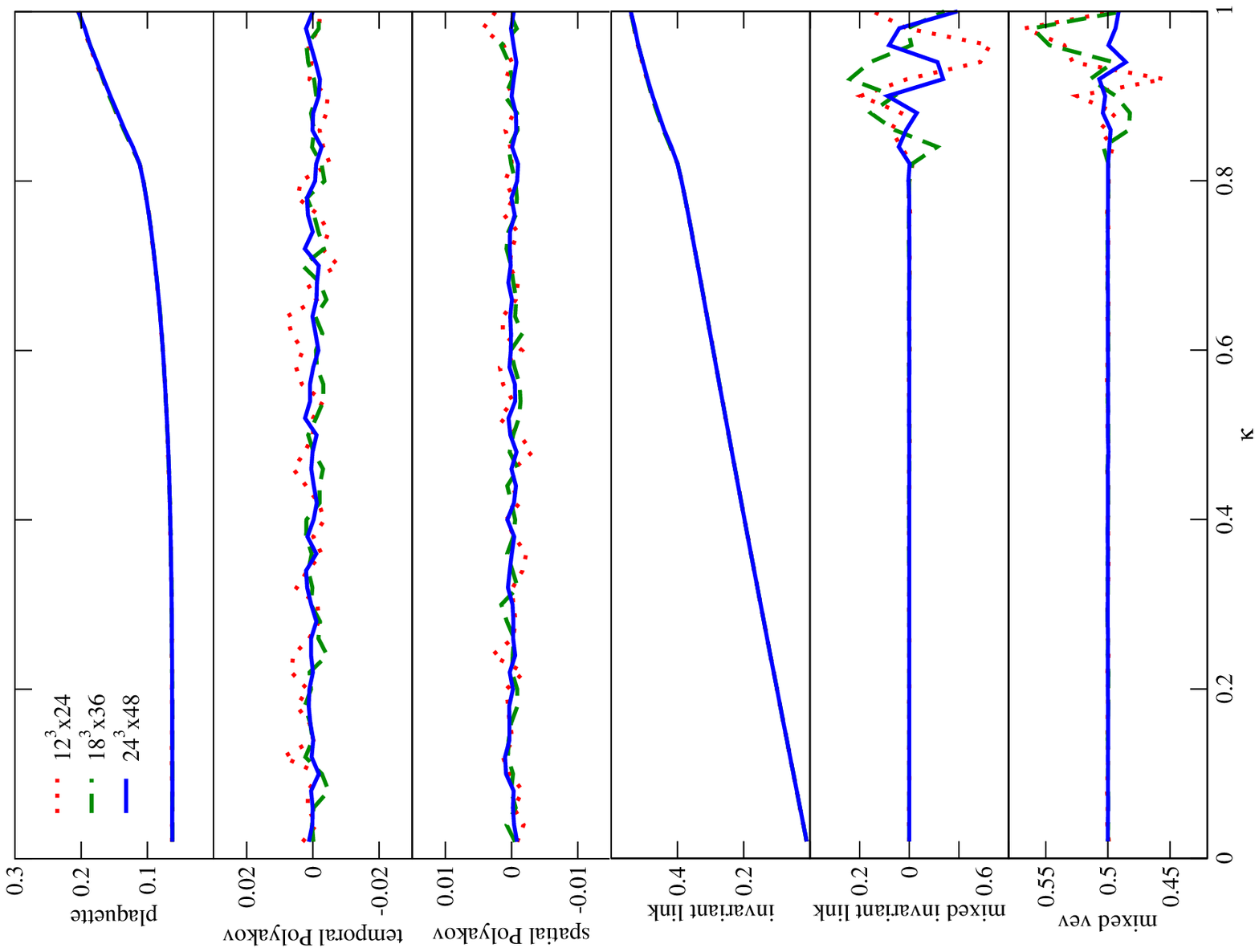}
\caption{Indications of the phase transition from a variety of observables
         for the fixed-length theory at $\beta=0.25$.
         Three different lattice sizes are shown.}
\label{fig_betap25}
\end{figure}

Phase transitions are readily identified, on a lattice with $N$ sites,
by scanning through parameter
space and computing simple observables such as the average plaquette
\begin{equation}
\frac{1}{2N}\sum_{x,\mu<\nu}{\rm Tr}U_\mu(x)U_\nu(x+\mu)U^\dagger_\mu(x+\nu)
U^\dagger_\nu(x) \,,
\end{equation}
Polyakov loops
\begin{equation}
\frac{1}{2N}\sum_x{\rm Tr}\prod_nU_4(x+n\hat 4) \,,
\end{equation}
gauge-invariant links, where $(i,j)$ = (1,1) or (1,2) or (2,2),
\begin{equation}\label{gaugeinvariantlinks}
L_{ij} \equiv \frac{1}{N}\sum_x\left(\Phi_i^\dagger(x)U_\mu(x)\Phi_j(x+\mu)
 + {\rm h.c.}\right) \,,
\end{equation}
and the mixed vev
\begin{equation}
\frac{1}{N}\sum_x\left|\Phi_1^\dagger(x)\Phi_2(x)\right|^2 \,.
\end{equation}
Note that the mixed gauge-invariant link ($i\neq j$) and mixed vev do not
preserve the
intradoublet symmetries, and recall that Polyakov loops are order
parameters for confinement in the pure gauge theory and are sometimes
used to provide a nonrigorous definition of confinement in the one-doublet
SU(2)-Higgs model\cite{Greensite:2006ng}.
Figs.~\ref{fig_beta4} and \ref{fig_betap25}
show examples of scanning through $\kappa_1=\kappa_2$ values
while holding $\beta$ fixed in the fixed-length theory.
For the mixed invariant link and mixed vev, statistical errors (not shown)
scale inversely with $\sqrt{N}\sqrt{\#\rm configurations}$ at small $\kappa$ but
they scale inversely with $\sqrt{\#\rm configurations}$ at large $\kappa$.
This behavior is indicative of the spontaneous breaking of intradoublet
symmetries.
The Polyakov loops are affected by the phase transition at $\beta=4$ but not
at $\beta=0.25$, suggesting that the phase transition separates a
confinement region from a Higgs region at large $\beta$ only.  This is
precisely how the confinement/Higgs transition melts away in the one-doublet
SU(2)-Higgs model as well\cite{Fradkin:1978dv}.

Figs.~\ref{fig_beta4} and \ref{fig_betap25} also indicate that the
location of the phase transition is rather insensitive to the size of the
lattices employed.
The average plaquette and gauge-invariant link undergo a qualitative
change at the phase transition for all $\beta$, and in practice the
gauge-invariant link is a
convenient first diagnostic when searching for the phase transition.

\begin{figure}
\includegraphics[width=11cm,angle=270,trim=65 0 0 350,clip=true]
{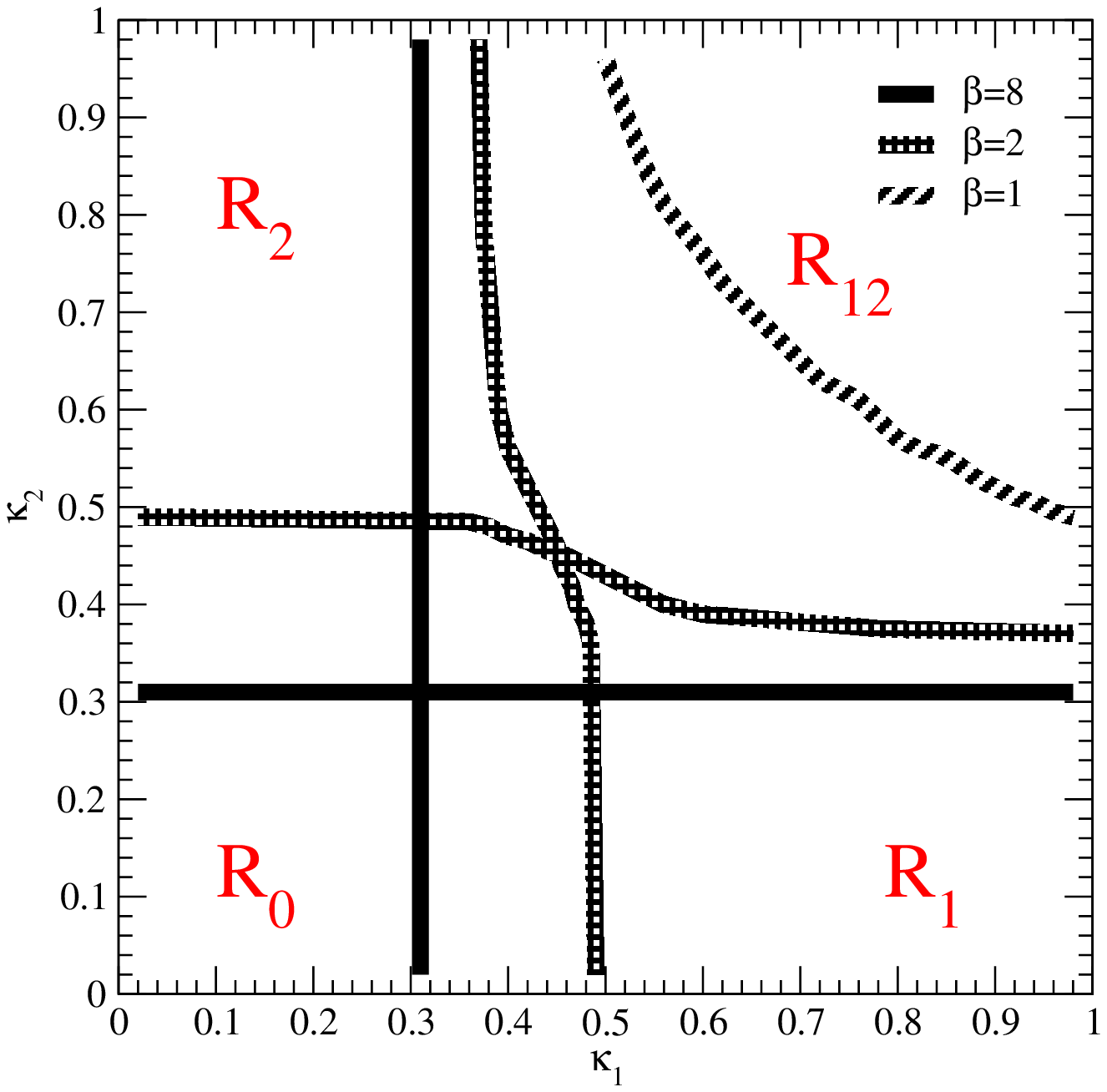}
\caption{Phase diagram for the fixed-length theory as computed for three
         different $\beta$ values on $16^4$ lattices.
         $R_{12}$ is the broken-symmetry region where $L_{12}$ of
         Eq.~(\protect\ref{gaugeinvariantlinks}) is nonzero.  All other regions
         are analytically connected at small $\beta$ but quantitatively
         distinguished at larger $\beta$ by finding large $L_{11}$ values in
         region $R_1$, large $L_{22}$ values in region $R_2$, and small
         values for both $L_{11}$ and $L_{22}$ in region $R_0$.}
\label{fig_PhaseDiagram1}
\end{figure}

\begin{figure}
\includegraphics[width=11cm,angle=270,trim=65 0 0 350,clip=true]
{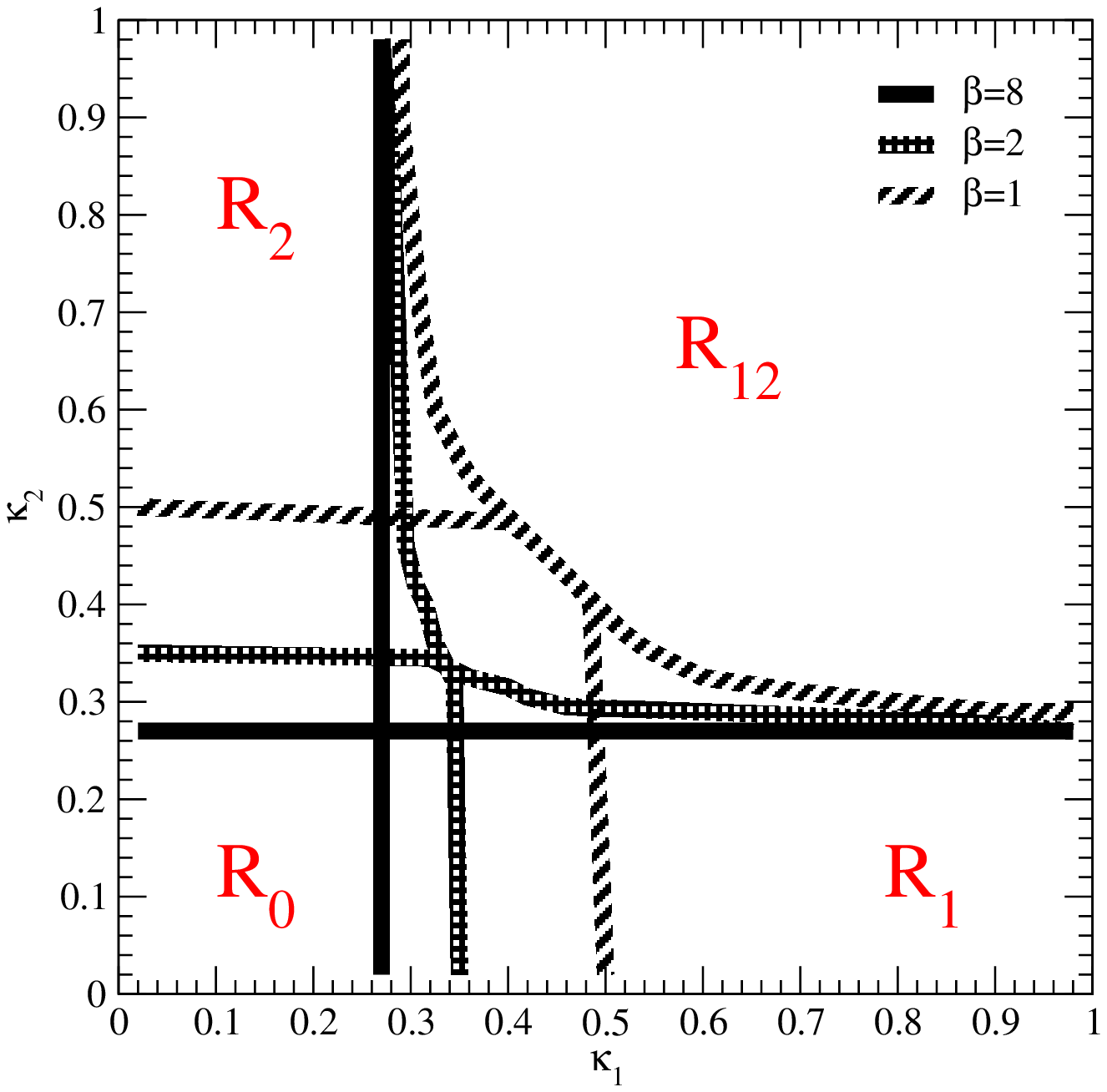}
\caption{Phase diagram for the theory with $\lambda_1=\lambda_2=1$ and
         $\lambda_{12}=0$, as computed for three
         different $\beta$ values on $16^4$ lattices.
         $R_{12}$ is the broken-symmetry region where $L_{12}$ of
         Eq.~(\protect\ref{gaugeinvariantlinks}) is nonzero, region $R_1$ has
         large $L_{11}$, region $R_2$ has large $L_{22}$, and region $R_0$ has
         small $L_{11}$ and $L_{22}$.}
\label{fig_PhaseDiagram2}
\end{figure}

The phase diagram for the fixed-length theory is shown for three different
$\beta$ values in Fig.~\ref{fig_PhaseDiagram1}.
Since large $\beta$ corresponds to weak gauge coupling, it is not surprising
that one finds two orthogonal phase transitions: one separating the
Higgs and confinement phases of the first scalar field (and therefore
essentially independent of $\kappa_2$ in the figure) and the other for the
second scalar field (essentially independent of $\kappa_1$).
This divides the $\kappa_1,\kappa_2$ plane into four regions but these are
not four separate phases as is evident from results at smaller $\beta$.
At $\beta=2$ the phase transitions affect one another near their
mutual crossing point, and for $\beta=1$
only a single phase transition is evident.
The corresponding data for $\lambda_1=\lambda_2=1$, with $\lambda_{12}=0$,
are given in Fig.~\ref{fig_PhaseDiagram2}.

\begin{figure}
\includegraphics[width=15cm,angle=270,trim=20 350 80 50,clip=true]{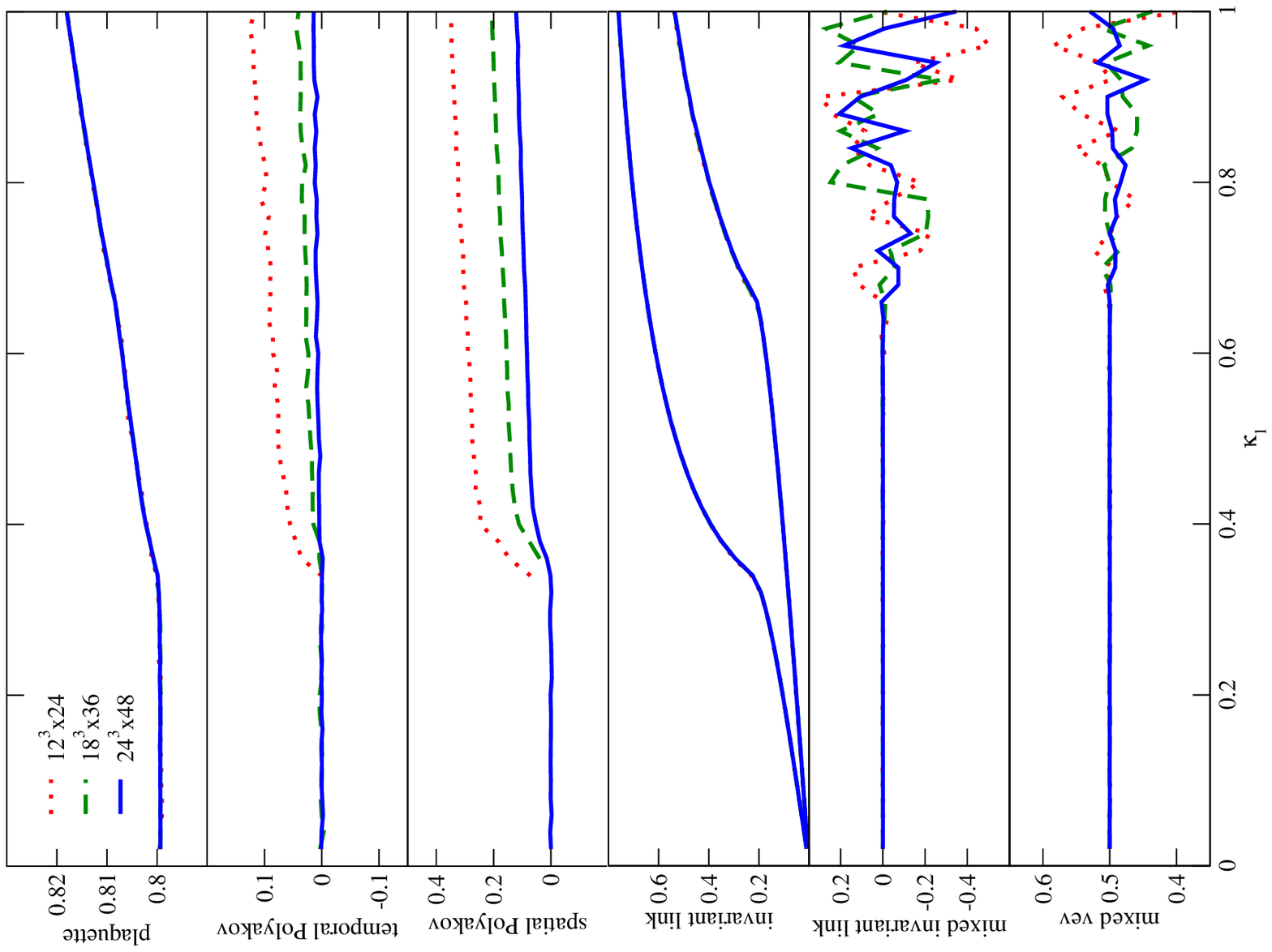}
\caption{Indications of the phase transition from a variety of observables
         for the fixed-length theory at $\beta=4.0$ with $\kappa_1=2\kappa_2$.
         Three different lattice sizes are shown.}
\label{fig_fs1p1beta4pub}
\end{figure}

\begin{figure}
\includegraphics[width=15cm,angle=270,trim=20 350 80 50,clip=true]{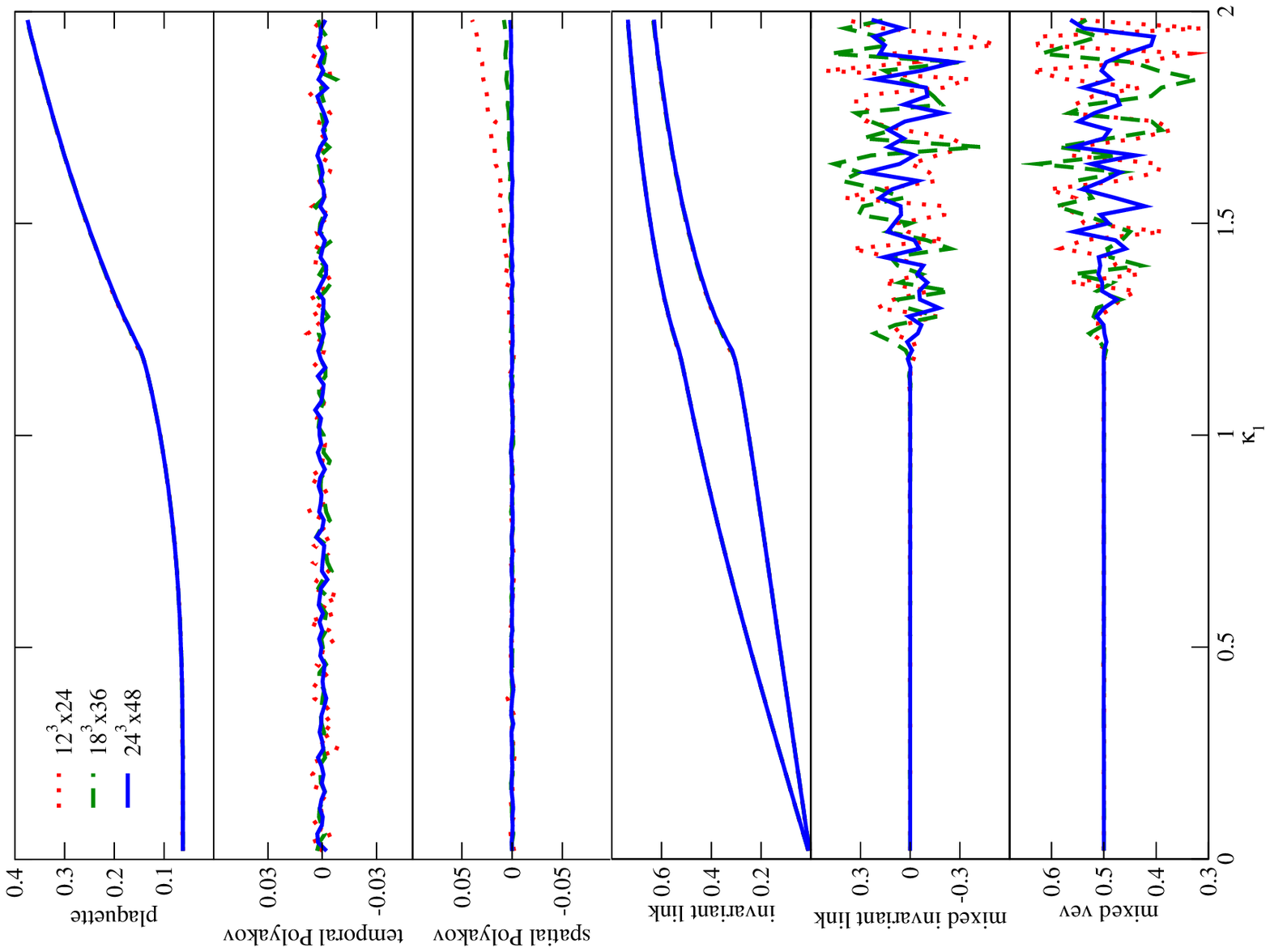}
\caption{Indications of the phase transition from a variety of observables
         for the fixed-length theory at $\beta=0.25$ with $\kappa_1=2\kappa_2$.
         Three different lattice sizes are shown.}
\label{fig_fs1p1betap25pub}
\end{figure}

As is clear from Figs.~\ref{fig_PhaseDiagram1} and \ref{fig_PhaseDiagram2},
there is always a single phase
transition in a theory with degenerate scalar fields ($\kappa_1=\kappa_2$)
but nondegenerate fields typically have more.  Choosing $\kappa_1=2\kappa_2$
for definiteness, it is not clear from Fig.~\ref{fig_PhaseDiagram1}
which observables display a qualitative change at which phase boundaries, so
this information is provided in Figs.~\ref{fig_fs1p1beta4pub} and
\ref{fig_fs1p1betap25pub}.
The Polyakov loops are zero in the $R_0$ region
but with no clear transition for small $\beta$,
while the observables that mix $\Phi_1$ and $\Phi_2$ display their
qualitative change at the $R_{12}$ boundary.
These results suggest that the $R_0$ region
be viewed as the confinement region,
and the $R_{12}$ region is the
phase of broken intradoublet symmetry.

\begin{figure}
\includegraphics[width=115mm,angle=270,trim=65 0 0 0,clip=true]
{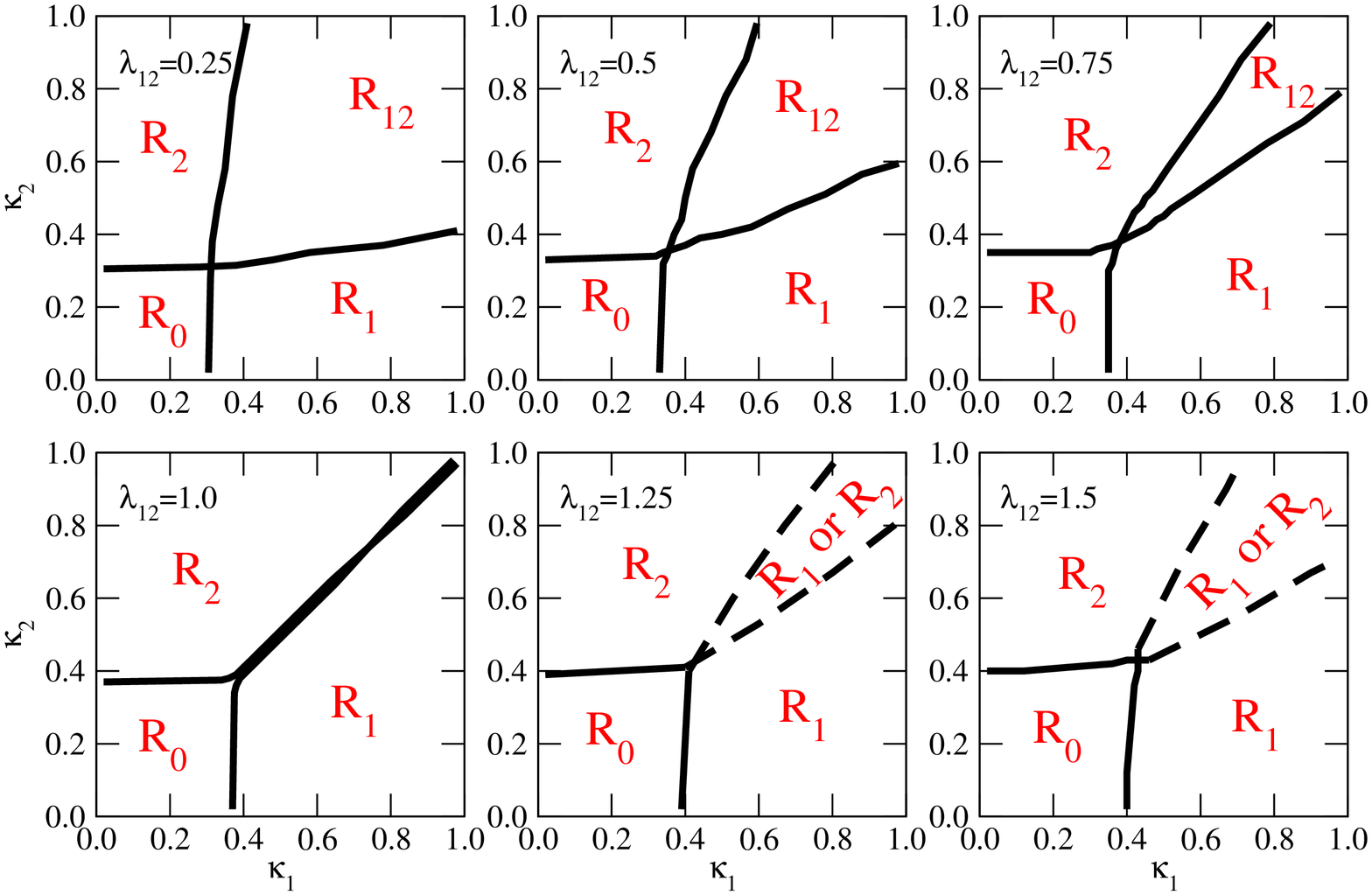}
\caption{Phase diagram for the theory with $\lambda_1=\lambda_2=1$
         at $\beta=8.0$, as computed for six different $\lambda_{12}$ values.
         $R_{12}$ is the broken-symmetry region where $L_{12}$ of
         Eq.~(\protect\ref{gaugeinvariantlinks}) is nonzero, region $R_1$ has
         large $L_{11}$, region $R_2$ has large $L_{22}$, and region $R_0$ has
         small $L_{11}$ and $L_{22}$.  For $\lambda_{12}>1$, hysteresis is
         observed in lieu of symmetry breaking.}
\label{fig_PhaseDiagramsXi8pub}
\end{figure}

All of the simulations discussed so far have used $\lambda_{12}=0$, but
it is interesting to explore nonzero values of this parameter since
the lattice action has an interdoublet symmetry when
($\kappa_1$=$\kappa_2$, $\lambda_1$=$\lambda_2$=$\lambda_{12}$).
The effect on the phase diagram due to variation of $\lambda_{12}$ is
plotted in Fig.~\ref{fig_PhaseDiagramsXi8pub}
for the case of $\beta=8.0$, $\lambda_1=\lambda_2=1$.
The two phase transition lines, which were essentially straight and
orthogonal at $\lambda_{12}=0$ in Fig.~\ref{fig_PhaseDiagram2}, bend toward
one another at large hopping parameters as $\lambda_{12}$ is increased.
This pinching of the phase of broken intradoublet symmetry continues until
that phase is reduced to a single line at $\lambda_{12}=1$.  That line runs
along $\kappa_1=\kappa_2$ which is precisely where the extra interdoublet
symmetry is manifest in the lattice action.

As $\lambda_{12}$ is increased beyond 1, the phase of broken intradoublet
symmetry vanishes and a region of hysteresis emerges, bounded in
Fig.~\ref{fig_PhaseDiagramsXi8pub}
by dashed lines.  Only one, not both, of the scalar fields is in
its Higgs phase in the region between the dashed lines, meaning that the
phenomenology of this region is similar to either the
$R_1$ region or the $R_2$ region.
Which of these options is realized between the dashed lines depends upon how
the dynamical system enters the region.  For example, if $\kappa_1$ is
gradually increased to pass through that region, then there will be no
qualitiative change in our standard suite of observables as the system
enters the region, but there will be a qualitative change as the system
exists from the region (by crossing the second dashed line).

\section{Spontaneous symmetry breaking}\label{sec_breakingsGoldstones}

\subsection{Qualitative features}

The simulations discussed above
found large fluctuations (for sufficiently large $\kappa_1$ and $\kappa_2$
values) for the ensemble averages of
Re($\Phi_1^\dagger(x)\Phi_2(x)$), Im($\Phi_1^\dagger(x)\Phi_2(x)$),
Re($\Phi_1^\dagger(x)\Phi_{c2}(x)$) and Im($\Phi_1^\dagger(x)\Phi_{c2}(x)$).
The sum of the squares of these four quantities is observed to have small
fluctuations for all $\kappa_1,\kappa_2$ values, and is close to zero for
small $\kappa_1,\kappa_2$ but far from zero for large $\kappa_1,\kappa_2$.

The large fluctuations become smaller when an explicit symmetry-breaking term,
such as
\begin{equation}\label{explicitbreaking}
\delta{\cal L} = \frac{\eta}{2}{\rm Tr}\left(\varphi_1^\dagger(x)\varphi_2(x)
                 \right) \,,
\end{equation}
is added to the theory.  Simulations can be performed for various values of
$\eta$ and then extrapolated to $\eta=0$.
For nonzero $\eta$, one finds $\left<{\rm Re}(\Phi_1^\dagger(x)\Phi_2(x))
\right>\neq0$ but the other three ensemble averages are statistically zero,
and $\left<{\rm Re}(\Phi_1^\dagger(x)\Phi_2(x))\right>$ itself approaches
zero as $\eta\to0$.

To determine which symmetry-group generators are broken, consider how these
ensemble averages are affected by the general symmetry transformation of
Eqs.~(\ref{generalPhiTransformation}) and (\ref{generalPhicTransformation}).
In particular, the use of
\begin{equation}
\left<{\rm Im}(\Phi_1^\dagger\Phi_2)\right>=
\left<{\rm Re}(\Phi_1^\dagger\Phi_{c2})\right>=
\left<{\rm Im}(\Phi_1^\dagger\Phi_{c2})\right>=0
\end{equation}
leads to
\begin{eqnarray}
\left<\Phi_1^\dagger\Phi_2\right>
&\to& \left(\cos\gamma_1\cos\gamma_2e^{i(\alpha_2-\alpha_1)}
      +\sin\gamma_1\sin\gamma_2e^{i(\beta_2-\beta_1)}\right)
      \left<{\rm Re}\left(\Phi_1^\dagger\Phi_2\right)\right> \,,
      \label{Phi1Phi2B} \\
\left<\Phi_1^\dagger\Phi_{c2}\right>
&\to& e^{-2i\theta}\sin\gamma_{12}\cos\gamma_{12}\left<\Phi_2^\dagger\Phi_2
    -\Phi_1^\dagger\Phi_1\right> \nonumber \\
&&  + e^{-2i\theta}\left[-\sin^2\gamma_{12}\sin\gamma_1\cos\gamma_2
      e^{i(\alpha_2+\beta_1)}-\cos^2\gamma_{12}\cos\gamma_1\sin\gamma_2
      e^{-i(\alpha_1+\beta_2)}\right. \nonumber \\
&&  + \left.\sin^2\gamma_{12}\cos\gamma_1\sin\gamma_2
      e^{i(\alpha_1+\beta_2)}+\cos^2\gamma_{12}\sin\gamma_1\cos\gamma_2
      e^{-i(\alpha_2+\beta_1)}\right]\left<{\rm Re}\left(\Phi_1^\dagger\Phi_2
      \right)\right> \,. \nonumber \\ \label{Phi1Phic2B}
\end{eqnarray}
Begin with the situation where the original global symmetry
was only SU(2)$\times$SU(2).  As noted in the previous section, this
corresponds to $\gamma_{12}=0$.  Therefore Eqs.~(\ref{Phi1Phi2B}) and
(\ref{Phi1Phic2B}) are simply
\begin{eqnarray}
\left<\Phi_1^\dagger\Phi_2\right>
&\to& \left(\cos\gamma_1\cos\gamma_2e^{i(\alpha_2-\alpha_1)}
      +\sin\gamma_1\sin\gamma_2e^{i(\beta_2-\beta_1)}\right)
      \left<{\rm Re}\left(\Phi_1^\dagger\Phi_2\right)\right> \,, \\
\left<\Phi_1^\dagger\Phi_{c2}\right> &\to& 0 \,.
\end{eqnarray}
The maximal unbroken subgroup is obtained from
the case of $(\alpha_1,\beta_1,\gamma_1)=(\alpha_2,\beta_2,\gamma_2)$, which
identifies a residual global SU(2).  Therefore
\begin{equation}
{\rm SU(2)}\times{\rm SU(2)}
~~\to~~
{\rm SU(2)}
~~~~~{\rm if}~ (\kappa_1=\kappa_2,\lambda_1=\lambda_2=\lambda_{12})
{\rm ~is~not~true.}
\end{equation}

Now consider Eqs.~(\ref{Phi1Phi2B}) and (\ref{Phi1Phic2B}) in the special case
where the original global symmetry was
SU(2)$\times$SU(2)$\times$U(1)$\times$U(1).
Now $\gamma_{12}\neq0$, but again
the largest unbroken subgroup is obtained from
the case of $(\alpha_1,\beta_1,\gamma_1)=(\alpha_2,\beta_2,\gamma_2)$, which
gives
\begin{eqnarray}
\left<\Phi_1^\dagger\Phi_2\right>
&\to& \left<{\rm Re}\left(\Phi_1^\dagger\Phi_2\right)\right> \,, \\
\left<\Phi_1^\dagger\Phi_{c2}\right>
&\to& e^{-2i\theta}\sin\gamma_{12}\cos\gamma_{12}\left<\Phi_2^\dagger\Phi_2
    -\Phi_1^\dagger\Phi_1\right> \,.
\end{eqnarray}
As will be discussed below,
lattice simulations find $\left<\Phi_2^\dagger\Phi_2\right>
\neq\left<\Phi_1^\dagger\Phi_1\right>$.  Therefore $\gamma_{12}=n\pi/2$ for
some integer $n$, and $\theta$ remains as a symmetry generator in the theory.
As a consequence, the symmetry breaking in this special case is 
\begin{equation}\label{extrabrokenU1}
{\rm SU(2)}\times{\rm SU(2)}\times{\rm U(1)}\times{\rm U(1)}
~~\to~~
{\rm SU(2)}\times{\rm U(1)}
~~~~~{\rm if}~ (\kappa_1=\kappa_2,\lambda_1=\lambda_2=\lambda_{12})
{\rm ~is~true.}
\end{equation}

\subsection{Symmetry-breaking order parameter}

The calculations of the previous section can give qualitative information
about the phase diagram, but getting a quantitative estimate of the
order parameter in the broken phase requires a different approach.
The system has to be forced to choose between different degenerate
vacua by
inserting an explicit symmetry-breaking term such as
Eq.~(\ref{explicitbreaking}) into the theory and then studying
the limit as the coefficient, $\eta$, approaches zero. The infinite-volume
limit should be taken before removing the symmetry-breaking term.

\begin{figure}
\includegraphics[width=7cm,angle=270,trim=80 20 200 0,clip=true]{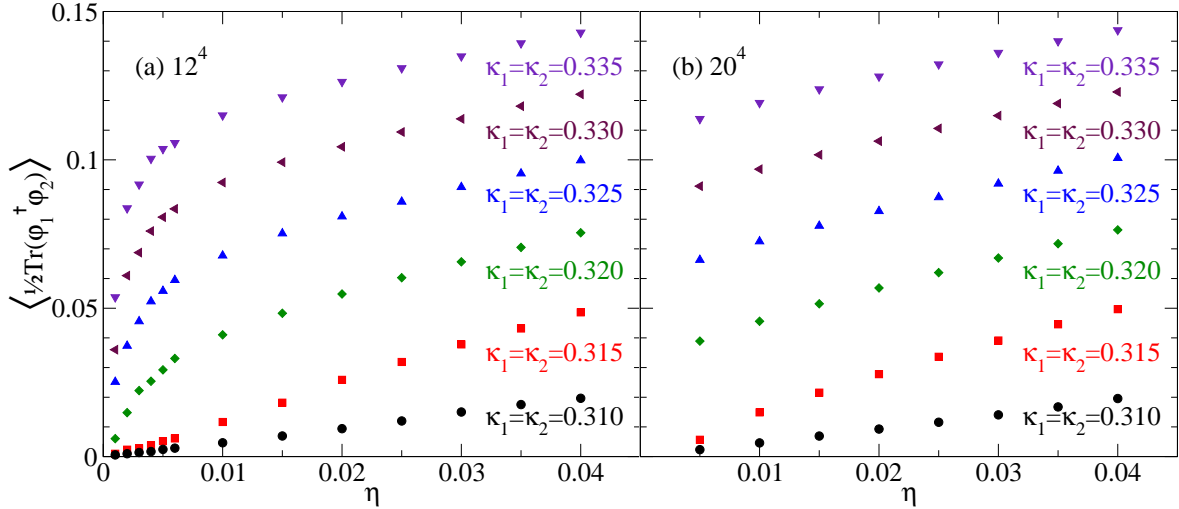}
\caption{A vacuum expectation value that breaks SU(2)$\times$SU(2) is
graphed as a function of its Lagrangian coefficient.  These data are from
simulations for the fixed-length theory with $\beta=8$ on
(a) $12^4$ and (b) $20^4$ lattices.}
\label{fig_etapub}
\end{figure}

Results of numerical simulations on finite size ($12^4$ and $20^4$) lattices
are displayed in Fig.~\ref{fig_etapub}.
For hopping parameters below the phase transition (i.e.\ $\kappa<\kappa_c$),
the symmetry-breaking
vev extrapolates linearly to zero as $\eta$ vanishes.
For hopping parameters above the phase transition ($\kappa>\kappa_c$),
the symmetry-breaking
vev appears to extrapolate to nonzero values, except for a bending
toward zero at small $\eta$ (visible in the $12^4$ simulation).   
This decrease reflects the fact that there is no true spontaneous symmetry 
breaking in a finite system  and is due to modes whose Compton wavelength 
becomes larger than the lattice size at small $\eta$. As the volume is 
increased this effect is restricted to a smaller region near $\eta$ = 0.

To deal with the volume effect, we recall that the effective
field theory in finite volume for a scalar theory in the broken phase 
was developed long ago\cite{Hasenfratz:1989pk,Hasenfratz:1990fu}
and was studied numerically in some detail for the one-doublet
model\cite{Hasenfratz:1988kr}.

\begin{figure}
\includegraphics[width=8cm,angle=270,trim=140 0 100 70,clip=true]{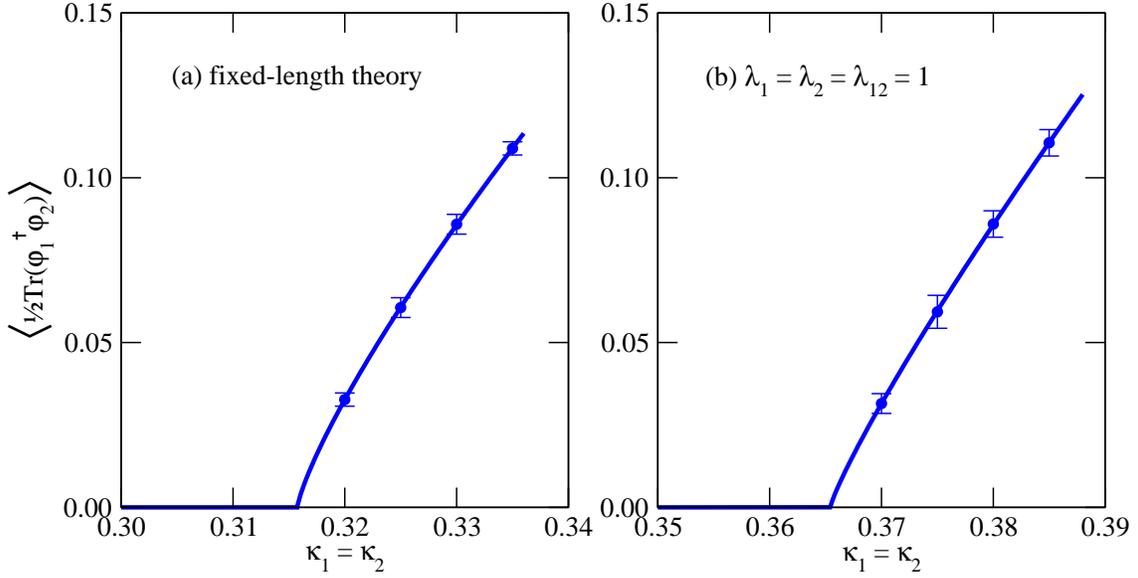}
\caption{An order parameter for SU(2)$\times$SU(2) breaking after
extrapolation to $\eta=0$.  These data are from
simulations with $\beta=8$.}
\label{fig_vevpub}
\end{figure}

Doing calculations of $\left<\frac{1}{2}{\rm Tr}\varphi_1^\dagger\varphi_2\right>$
for different volumes (from $8^4$ to $20^4$) and  different $\eta$ and using 
procedures which are verified by study of the one-doublet model,
we can estimate the infinite-volume value of the order parameter.
Figure~\ref{fig_vevpub}
shows the value of the order parameter for the SU(2)$\times$SU(2)
symmetry breaking in two cases: the fixed-length theory and the theory with
$\lambda_1=\lambda_2=\lambda_{12}=1$.  Using a parametrization
$\left<\frac{1}{2}{\rm Tr}\varphi_1^\dagger\varphi_2\right>\propto(\kappa-\kappa_{c})^{\nu}$
for $\kappa>\kappa_{c}$ we can estimate the critical $\kappa$. The values
of $\kappa_{c}$ corresponding to the lines in Fig.~\ref{fig_vevpub} are 0.316
for the fixed-length theory and 0.365 for $\lambda_1=\lambda_2=\lambda_{12}=1$.

\begin{figure}
\includegraphics[width=10cm,angle=270,trim=85 0 0 70,clip=true]{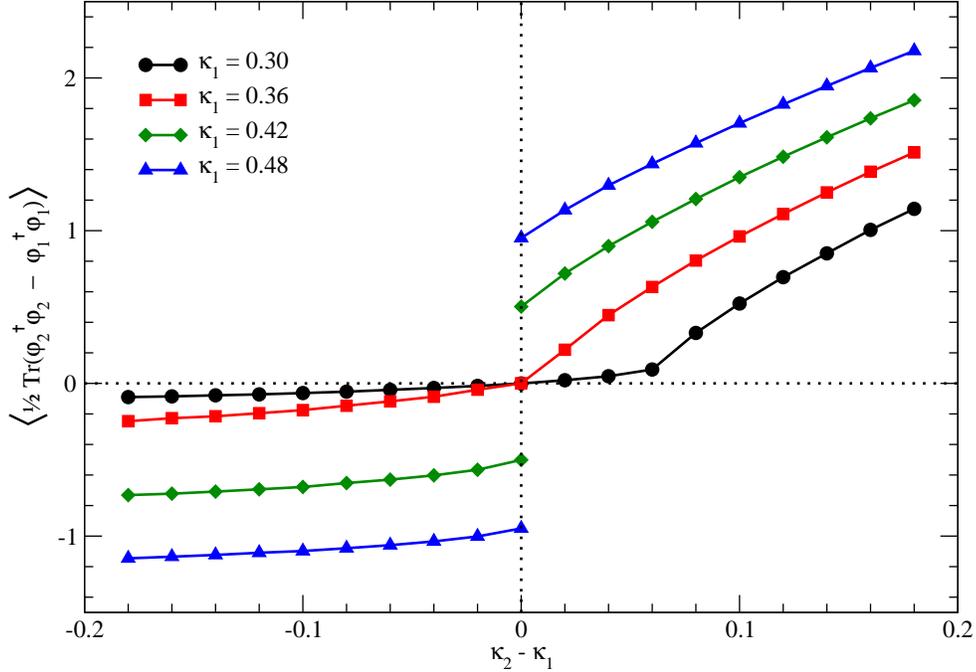}
\caption{An order parameter for spontaneous breaking of the extra U(1).
These data are from
simulations with $\beta=8$ and $\lambda_1=\lambda_2=\lambda_{12}=1$.}
\label{fig_extraU1pub}
\end{figure}

Recall the prediction from Eq.~(\ref{extrabrokenU1}) of an extra broken U(1)
symmetry in the theory when $\kappa_1=\kappa_2$ and
$\lambda_1=\lambda_2=\lambda_{12}$.
This is verified by adding a symmetry-breaking term to the theory, and then
extrapolating its coefficient to zero.
In fact, the simplest way to add an
appropriate extra term is to run simulations with
$\kappa_1\neq\kappa_2$ and extrapolate the results to $\kappa_1=\kappa_2$.
An example is provided in Fig.~\ref{fig_extraU1pub}.
The transition from broken to unbroken U(1) is found to occur at the same
critical hopping parameter as the breaking of the SU(2)$\times$SU(2).
At first glance, the kink in the $\kappa_1=0.48$ curve of
Fig.~\ref{fig_extraU1pub}
may be puzzling, but comparison to Fig.~\ref{fig_PhaseDiagramsXi8pub}
makes the interpretation clear: the kink occurs at the phase transition
crossed by varying $\kappa_2$ while holding $\kappa_1$ fixed.

\subsection{Goldstone bosons}

\begin{figure}
\includegraphics[width=10cm,angle=270,trim=85 0 0 70,clip=true]
{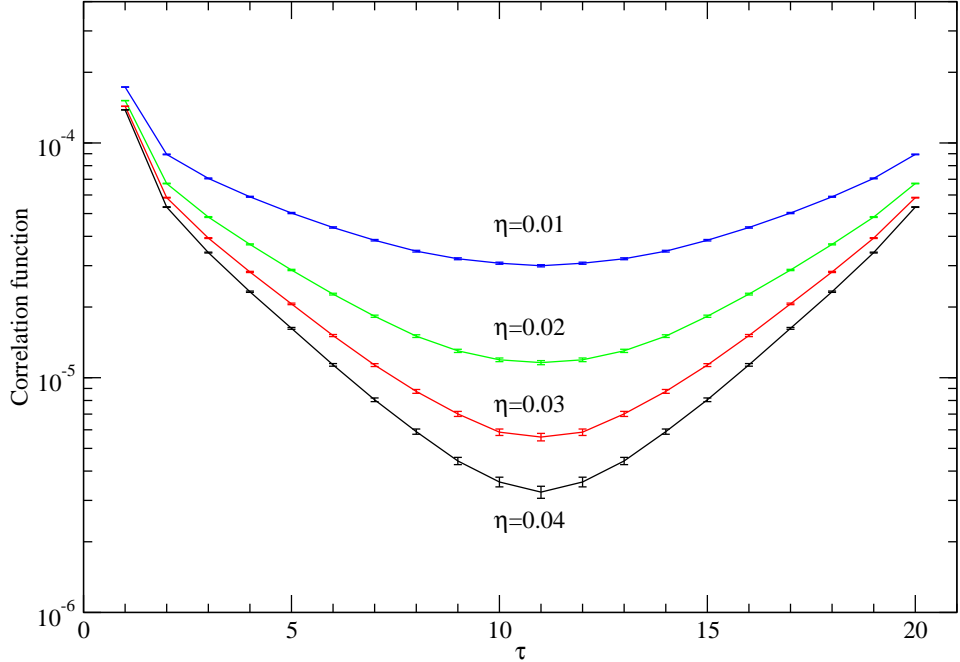}
\caption{Correlation functions for the operators defined in
Eq.~(\protect\ref{GBoperators}).
These data are from simulations on $16^3\times20$ lattices
with $\beta=8$ in the fixed-length theory.}
\label{fig_cfscaLinfpub}
\end{figure}

\begin{figure}
\includegraphics[width=10cm,angle=270,trim=85 0 0 70,clip=true]
{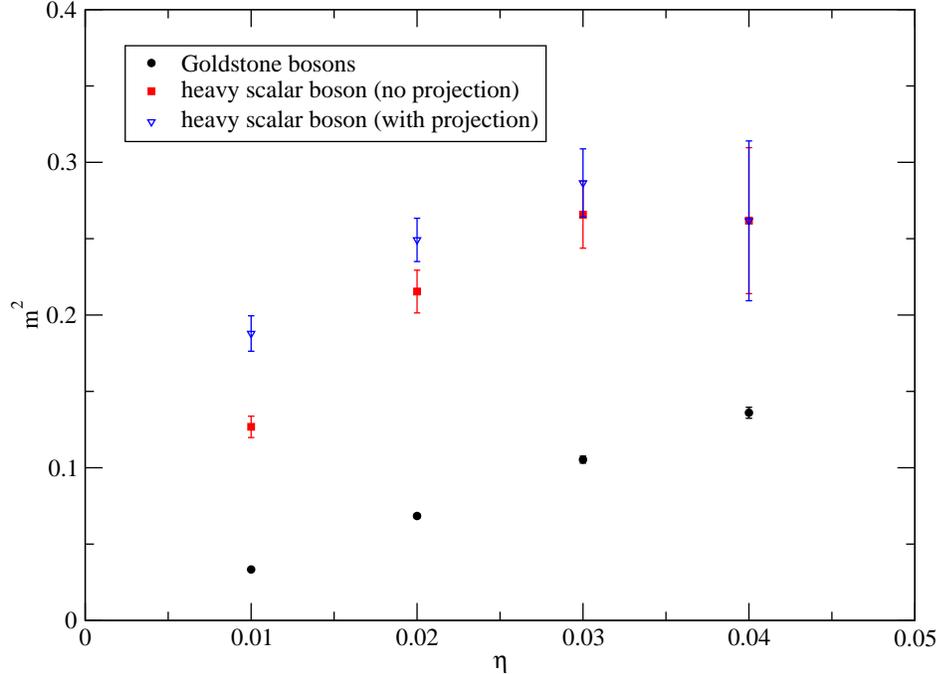}
\caption{Squared masses for the Goldstone bosons obtained from
Eq.~(\protect\ref{GBoperators}) and the scalar boson obtained from
Eq.~(\protect\ref{sigmaoperator}).
These data are from simulations on $16^3\times20$ lattices
with $\beta=8$ in the fixed-length theory.}
\label{fig_masssqLinfpub}
\end{figure}

Spontaneous breaking of any continuous global symmetry generates a
Goldstone boson for each broken generator.
The three Goldstone bosons arising from
SU(2)$\times$SU(2)$\rightarrow$SU(2) are found to couple readily to the
operators
\begin{equation}\label{GBoperators}
\frac{1}{2}{\rm Tr}\left[\varphi_1^\dagger\varphi_2\tau_a\right]
\end{equation}
where $\tau_1$, $\tau_2$ and $\tau_3$ are the standard Pauli matrices.
These operators are invariant under the
unbroken global SU(2) and under the gauge symmetry (which is never broken).
Examples of correlators for a range of $\eta$ are shown in
Fig.~\ref{fig_cfscaLinfpub} and the corresponding squared masses are shown in
Fig.~\ref{fig_masssqLinfpub}.  The small statistical errors provide convincing
evidence that the Goldstone boson squared mass vanishes with a linear
extrapolation of $\eta\to0$.  For comparison, the graph also contains
results for the operator
\begin{equation}\label{sigmaoperator}
\frac{1}{2}{\rm Tr}\left[\varphi_1^\dagger\varphi_2\right] \,.
\end{equation}
That operator is not an SU(2) triplet and does not couple to the Goldstone
bosons, but it does provide evidence of a heavy scalar particle in the
theory.  One might wish to name the Goldstone bosons $\pi_a$ and the extra
scalar boson ``$\sigma$'' to follow familiar notational conventions.
In addition to the direct method for obtaining the $\sigma$ correlation
function, the projection method described in \cite{Hasenfratz:1988kr}
was also used.  This projection is given by
\begin{equation}
{\cal O}_{\rm proj} = \sum_{a=1}^3\frac{M_a}{|M|}
\frac{1}{2}{\rm Tr}\left[\varphi_1^\dagger\varphi_2\tau_a\right]
\end{equation}
where, on a lattice with $N$ sites,
\begin{eqnarray}
M_a &=& \frac{1}{N}\sum_x
\frac{1}{2}{\rm Tr}\left[\varphi_1^\dagger\varphi_2\tau_a\right] \,, \\
|M| &=& \sqrt{M_1^2+M_2^2+M_3^2} \,.
\end{eqnarray}

\begin{figure}
\includegraphics[width=10cm,angle=270,trim=85 0 0 70,clip=true]{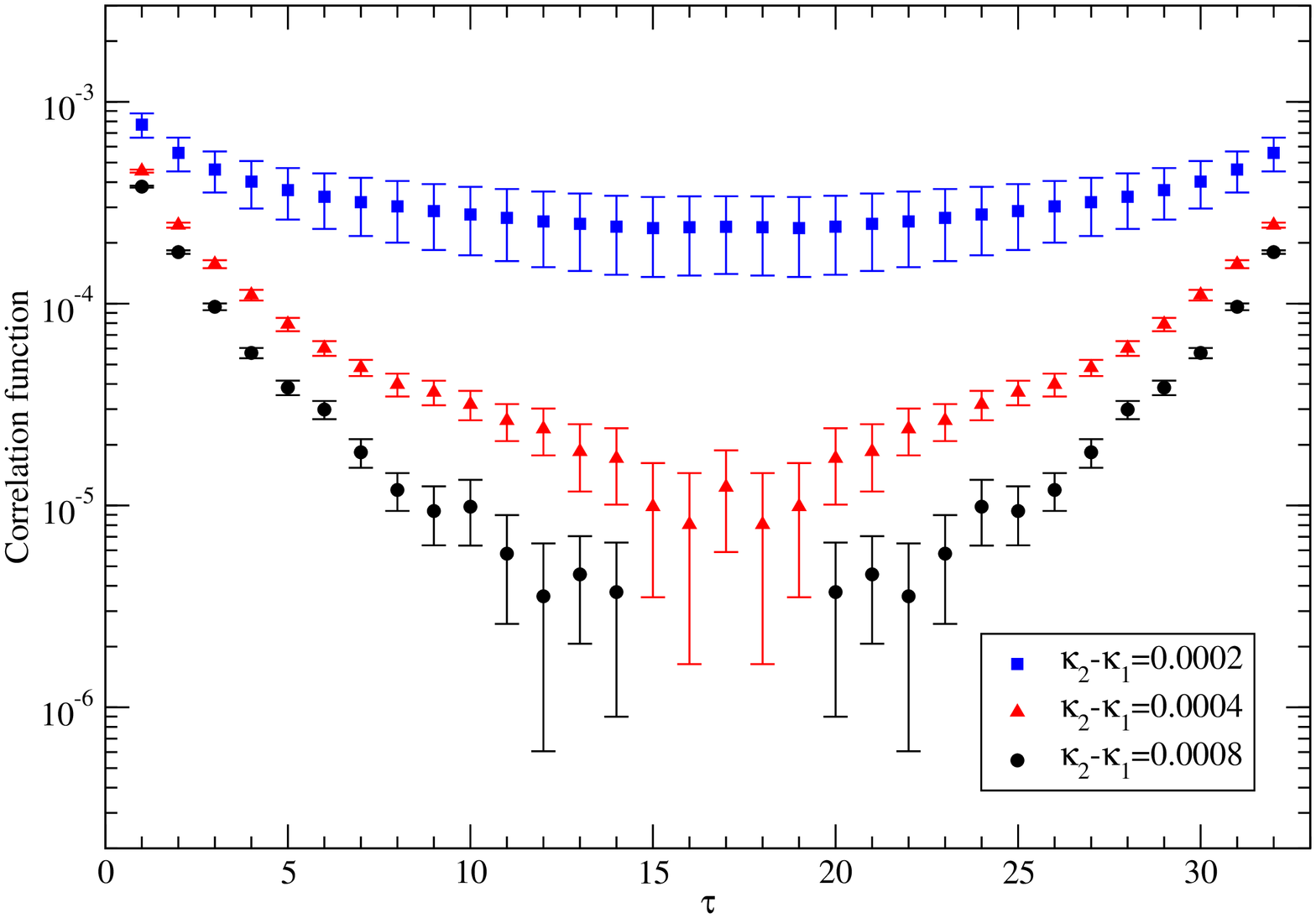}
\caption{Correlation functions for the operator obtained from
Eq.~(\protect\ref{u1operator}).
These data are from simulations on $16^3\times32$ lattices
with $\beta=2$, $\kappa_1=0.48$ and $\lambda_1=\lambda_2=\lambda_{12}=1$.}
\label{fig_u1corrpub}
\end{figure}

A possible operator for producing the Goldstone boson associated with the
breaking of an extra U(1) symmetry present when 
($\kappa_1$=$\kappa_2$, $\lambda_1$=$\lambda_2$=$\lambda_{12}$) is
\begin{equation}\label{u1operator}
\frac{1}{2}{\rm Tr}\left[\varphi_1^\dagger\varphi_1-\varphi_2^\dagger\varphi_2
\right] \,.
\end{equation}
Numerical simulations using this operator produced sizable
statistical fluctuations as shown in Fig.~\ref{fig_u1corrpub},
but are consistent with a mass that vanishes as $\eta\to0$.

\section{Summary}\label{sec_conclusion}

In this paper we studied an SU(2) Higgs model using lattice 
field theory methods. This approach provides a view of
symmetry breaking which is different from the one familiar
from the usual perturbative treatment of the standard model.
The difference stems from the use of the gauge field link in 
the lattice formulation. This removes the requirement of gauge
fixing and allows all quantities to be calculated in a 
gauge-invariant way. The physical states of the theory are
described by gauge-invariant operators which are necessarily
composite. The lattice Higgs model is in this sense not unlike QCD.

The SU(2) lattice Higgs model with one scalar doublet was studied long
ago. Regions of parameter space with seemingly different physical
behavior were identified by examining the scalar and vector
particle spectrum. These were associated with confined and 
Higgs ``phases''. However, it was suggested that in fact the model 
with only one fundamental scalar doublet has only one phase and 
no symmetries, local or global, are broken. Numerical simulations
are in accord with this expectation. The low-lying spectrum of the
theory consists of a massive scalar boson and a degenerate triplet
of massive vector bosons.

The addition of a second scalar doublet can lead to a richer 
symmetry structure than in the one-doublet model. For the model 
studied in this paper the global symmetry is generically 
SU(2)$\times$SU(2) but is enlarged to 
SU(2)$\times$SU(2)$\times$U(1)$\times$U(1) for particular 
parameter choices which allow for symmetry under interdoublet 
mixing. By examining the vacuum expectation values of a variety of 
operators, the phase diagram was mapped out. The confined and 
Higgs regions associated with the individual doublets could
be identified. When the hopping parameters are sufficiently 
large a new phase, in which there is a strong correlation 
of the two doublet fields and the global symmetry is 
spontaneously broken, emerges. 

The spontaneous symmetry breaking was verified by calculation
of the order parameter for some specific choices of the 
model parameters. This was done by the usual procedure of introducing
an explicit symmetry-breaking term and studying the behavior of 
the system as the volume of the simulation was increased and the
symmetry-breaking term was removed. The presence of Goldstone bosons
in the broken phase
was verified by calculation of the correlation functions of appropriate 
gauge-invariant interpolating operators. In addition to Goldstone bosons
we also find scalar states which remain massive in all phases.
As in the one-doublet model the gauge symmetry is unbroken. 

The focus of this work was spontaneous global symmetry breaking so the
question of the spectrum of vector bosons was not addressed.
In the nonperturbative lattice approach the vector bosons are
composite particles and are expected to be massive in all regions
of the phase diagram. This can be confirmed by a cursory examination
of the correlation functions of the vector operators. However, the 
quantitative determination of the mass is 
a difficult problem due to the plethora of operators
that can be constructed which would require careful study of 
operator mixing and also decays due to the presence of light 
(pseudo-)Goldstone bosons in the theory. Such a study
might give some information about the nature of the theory in different  
regions of the parameter space. Work on the
three-dimensional Higgs model \cite{Philipsen:1998} gives some 
insight into the difficulty and benefit of a spectrum calculation.

\acknowledgments
This work was supported in part by the Natural Sciences and 
Engineering Research Council of Canada, and by computing resources of
the Shared Hierarchical Academic Research Computing Network\cite{sharcnet}.


\end{document}